\documentclass{aastex631}

\newcommand{\kms}{\,\mathrm{km} \, \mathrm{s}^{-1}}

\newcommand{\Msolpc}{\, \mathrm{M}_{\sun} \, \mathrm{pc}^{-2}}

\newcommand{\alphavir}{\alpha_{\mathrm{vir}}}

\newcommand{\Mcloud}{M_{\mathrm{cloud}}}
\newcommand{\Rcloud}{R_{\mathrm{cloud}}}
\newcommand{\Sigmacloud}{\Sigma_{\mathrm{cloud}}}
\newcommand{\Msun}{\mathrm{M}_{\sun}}
\newcommand{\Zsun}{\mathrm{Z}_{\sun}}

\newcommand{\ftrap}{f_{\mathrm{trap}}}

\newcommand{\epsf}{\epsilon_{*,\mathrm{f}}}

\usepackage[flushleft]{threeparttable}
\usepackage{booktabs}
\usepackage{multirow}

\begin{document}

\title{The Interplay between the IMF and Star Formation Efficiency through Radiative Feedback at High Stellar Surface Densities}

\author[0000-0002-0311-2206]{Shyam H. Menon}
\affiliation{Department of Physics and Astronomy, Rutgers University, 136 Frelinghuysen Road, Piscataway, NJ 08854, USA}
\affiliation{Center for Computational Astrophysics, Flatiron Institute, 162 5th Avenue, New York, NY 10010, USA}

\author[0000-0002-0041-4356]{Lachlan Lancaster}
\thanks{Simons Fellow}
\affiliation{Department of Astronomy, Columbia University,  550 W 120th St, New York, NY 10025, USA}
\affiliation{Center for Computational Astrophysics, Flatiron Institute, 162 5th Avenue, New York, NY 10010, USA}

\author[0000-0001-5817-5944]{Blakesley Burkhart}
\affiliation{Department of Physics and Astronomy, Rutgers University, 136 Frelinghuysen Road, Piscataway, NJ 08854, USA}
\affiliation{Center for Computational Astrophysics, Flatiron Institute, 162 5th Avenue, New York, NY 10010, USA}

\author[0000-0002-6748-6821]{Rachel S. Somerville}
\affiliation{Center for Computational Astrophysics, Flatiron Institute, 162 5th Avenue, New York, NY 10010, USA}

\author[0000-0003-4174-0374]{Avishai Dekel}
\affiliation{Center for Astrophysics and Planetary Science, Racah Institute of Physics, The Hebrew University, Jerusalem, 91904, Israel}
\affiliation{Santa Cruz Institute for Particle Physics, University of California, Santa Cruz, CA 95064, USA}

\author[0000-0003-3893-854X]{Mark R. Krumholz}
\affiliation{Research School of Astronomy and Astrophysics, Australian National University, Canberra ACT 2611, Australia}
\affiliation{ARC Centre of Excellence for Astronomy in Three Dimensions (ASTRO3D), Canberra ACT 2611, Australia}



\begin{abstract}

The observed rest-UV luminosity function at cosmic dawn ($z \sim 8-14$) measured by JWST revealed an excess of UV-luminous galaxies relative to many pre-launch theoretical predictions. A high star-formation efficiency (SFE) and a top-heavy initial mass function (IMF) are among the mechanisms proposed for explaining this excess.
Although a top-heavy IMF has been proposed for its ability to increase the light-to-mass ratio (\(\Psi_{\mathrm{UV}}\)), the resulting enhanced radiative pressure from young stars could decrease the star formation efficiency (SFE), potentially driving galaxy luminosities back down. In this Letter, we use idealized radiation hydrodynamic simulations of star cluster formation to explore the effects of a top-heavy IMF on the SFE of clouds typical of the high pressure conditions found at these redshifts. We find that the SFE in star clusters with solar neighbourhood-like dust abundance decreases with increasingly top-heavy IMF's -- by $\sim 20 \%$ for an increase of factor 4 in $\Psi_{\mathrm{UV}}$, and by $50 \%$ for a factor $ \sim 10$ in $\Psi_{\mathrm{UV}}$. However, we find that an expected decrease in the dust-to-gas ratio ($\sim 0.01 \times \mathrm{Solar}$) at these redshifts can completely compensate for the enhanced light output. This leads to a (cloud-scale; $\sim 10 \, \mathrm{pc}$) SFE that is $\gtrsim 70\%$ even for a factor 10 increase in $\Psi_{\mathrm{UV}}$, implying that highly efficient star formation is unavoidable for high surface density and low metallicity conditions. Our results suggest that a top-heavy IMF, if present, likely coexists with efficient star formation in these galaxies.

\end{abstract}



\section{Introduction} \label{sec:intro}

Pre-supernova feedback via radiation, jets, and winds emitted by young stars have been recognized to play a pivotal role in regulating star formation and dictating the lifecycle of giant molecular clouds (GMCs) in galaxies \citep{Chevance_23,Jeffreson_2024,Burkhart_2024}. This feedback disrupts GMCs in order $\sim$ unity dynamical timescales via the energy and momentum they impart \citep{Krumholz_Matzner_2009,Fall_2010,Thompson_Krumholz_2016} and drive turbulent motions that could further provide support against collapse \citep[e.g.,][]{MacLow_Klessen_2004,Krumholz_2006,Federrath_2010,Menon_2020,MenonEtAl2021,Garcia_2020,Appel_2022}. Numerical simulations have demonstrated that this limits the integrated star formation efficiency ($\epsilon_* = M_{*}/M_{\mathrm{gas}}$) -- defined as ratio of the final stellar mass $M_*$ formed to the available gas mass in the parent molecular cloud $M_{\mathrm{gas}}$ -- to values $\lesssim 10\%$ in environments typical of the local Universe \citep{Raskutti_2016,Kim_2016,Geen_2016,Grudic_2018,Kim_2018,Burkhart_2018,He_2019,Kim_2021,Fukushima_2021,Grudic_2022,Lancaster_2021c}.

However, it has become increasingly evident that this is not the case for GMC's typical of high ISM pressure environments ($P/k_{\mathrm{B}} \gtrsim 10^8 \, \mathrm{K} \,\mathrm{cm}^{-3}$), for which both models \citep{Fall_2010,Thompson_Krumholz_2016} and numerical simulations \citep{Grudic_2018,Fukushima_2021,Lancaster_2021c,Menon_2022b,Menon_2023,Polak_2023} suggest efficiencies $\epsilon_* \gtrsim 80\%$ because the energy/momentum deposition rates of feedback in this regime is unable to counteract gravity. Such pressures correspond to GMCs with surface densities ($\Sigma \gtrsim \Sigma_{\mathrm{crit}} = 10^3 \, \Msolpc$) -- 2--3 orders of magnitude higher than typical of GMCs in the local Universe -- which are the likely sites of so-called super-star cluster formation \citep[SSCs; e.g.,][]{Mcgrady_2005,Zwart_2010,Turner_2015,LindaSmith_2020}; observational estimates seem to be consistent with a high value of $\epsilon_*$ for these conditions \citep{Turner_2017,LindaSmith_2020, Emig_2020, Villas_2020,Costa_2021,Hao_2022,Mckinney_2023,Sun_2024}. The environments that host these conditions are relatively rare in the local Universe -- limited to scenarios such as nuclear starbursts \citep[e.g.,][]{Leroy_2018,Emig_2020,Levy_2021}, merging luminous infrared-bright galaxies \citep[e.g.,][]{Johnson_2015,Finn_2019,Inami_2022} and localized starbursts in dwarf galaxies \citep[e.g.,][]{Ochsendorf_2017,Oey_2017,Turner_2017}. On the other hand, the higher densities, gas fractions, merger rates and accretion rates of galaxies at higher redshift suggest that high-pressure conditions are more commonly realized at these epochs; indeed, conditions observed in dusty starburst galaxies \citep{Casey_2014}, pre-quiescent massive compact galaxies \citep{Diamond-Stanic_2012,Rupke_2019}, and proto-globular cluster candidates resolved via gravitational lensing \citep{Vanzella_2022,Vanzella_2022b,Pascale_2023} reflect these conditions. 

It is therefore timely in this context that JWST has revealed that this dense, clumpy, and compact mode of star formation may well be ubiquitous in the reionization era, through the discovery of extremely blue, UV-luminous, compact galaxies at redshifts $z\gtrsim 10$ \citep{Finkelstein_2023a,Casey_2023,Mcleod_2023,Robertson_2023,Harikane_2023,Morishita_2023}. The observed sizes ($\lesssim 0.5 \, \mathrm{kpc}$) of these objects indicate stellar surface densities ($\Sigma \gtrsim 10^4$--$10^5 \, \Msolpc$) that are comparable to or possibly somewhat higher than those in local super star clusters \citep[see e.g. Fig. 6 of][]{Casey_2023}. Highly magnified regions of lensed fields reveal systems at $z\sim 8$--10 that are comprised of multiple dense, intensely star forming clusters, possibly representing the formation sites of present-day globular clusters \citep{Mowla_2024,Adamo_2024}.
The observed numbers of these bright $z\gtrsim 10$ galaxies are in excess of the predictions of nearly all pre-launch models of galaxy formation, including both semi-analytic models and numerical hydrodynamic simulations \citep[e.g.][]{Dayal_2017,Wilkins_2022,Kannan_thesan,Kannan22_mtng,Yung_2023,Hassan_2023}. Moreover, models almost uniformly predict a much more rapid \emph{evolution} of the number density of bright galaxies with redshift at these early epochs than the observations indicate \citep{Finkelstein_2023}. Some of the proposed solutions to this tension allude to possibly distinct conditions in star-forming clouds in and around these galaxies, resulting in higher star formation efficiency and/or weaker stellar feedback \citep[e.g.][]{Yung_2023,Williams_2024}, or to the possibility of a top-heavy IMF, which could lead to higher light-to-mass ratios \citep{Yung_2023,Inayoshi_2022,Harikane_2023}. 

For example, the feedback-free model \citep[FFB;][]{Dekel_2023} posits that when both the gas density\footnote{A high density is required by the model to accommodate i) short free-fall times that enable bursts of star formation free from the effects of strong stellar winds at sub-solar metallicities, and ii) short cooling times for efficient fragmentation into star-forming clumps.} and surface density in star-forming clouds are high enough ($n > n_{\rm crit} \sim 3\times 10^3 {\rm cm}^{-3}$ and $\Sigma > \Sigma_{\rm crit} \sim 3 \times 10^3 \Msolpc$), and the metallicity is low but not negligible ($Z \sim 0.01-0.1 \Zsun$ ), star formation occurs in a burst over a free-fall time of $\sim 1 \mathrm{Myr}$, prior to the onset of supernova feedback and with only weak effects from stellar winds and radiative feedback. This leads to globally efficient star formation in $z\gtrsim 10$ galaxies, many of which are expected to satisfy these conditions. 
\citet{Li_2023} show that this model produces predictions that are consistent with the JWST observations. 

On the other hand, several studies have suggested that the IMF could be top-heavy at these redshifts due to a higher cosmic microwave background (CMB) temperature \citep{Chon_2022}, low metallicities \citep{Sharda_2022,Sharda_2023,Chon_2023} and/or the contribution of Population III stellar populations for which there is general agreement on the possibility of top-heaviness \citep{Larson_1998,OmukaiEtAl2005,Klessen_2023,Harikane_2023}. The associated higher UV luminosity per unit-mass from a top-heavy IMF could help reconcile the UV luminosity functions without requiring a high star formation efficiency \citep[e.g.,][]{Inayoshi_2022}. Indeed, \citet{Yung_2023} show that their fiducial (without changing $\epsilon_*$) semi-analytic model (SAM) can reproduce the observed UV luminosity function at $z\sim 11$ when they increase the UV luminosity-to-mass by a factor $\sim 4$. 

However, $\epsilon_*$ and the IMF are not necessarily independent of each other -- a top-heavy IMF and the associated increased level of radiative and wind feedback due to a higher fraction of massive stars (that dominate these modes of feedback) is very likely to affect $\epsilon_*$. The metallicity could also affect $\epsilon_*$ through its impacts on the dust abundance and cooling physics. Quantifying the interdependence of $\epsilon_*$ with the IMF/metallicity is crucial to shedding light on potential solutions to these surprising findings. This is also relevant in the context of star formation in extreme environments at lower redshifts, where regions of high surface density seem to show possible evidence of top-heavy IMFs \citep{Zhang_2018_IMF,Schneider_2018,Upadhyaya_2024}. While several previous authors have studied the impact of a top-heavy IMF on star cluster formation \citep{Fukushima_2023,Chon_2023}, they focus on clouds with mass surface densities and escape speeds that are lower ($\lesssim 10^3 \, \Msolpc$; $v_{\mathrm{esc}} \lesssim 20 \, \mathrm{km} \, \mathrm{s}^{-1}$) than the extreme cases being found with JWST. These studies have also been done use using numerical methods with less accurate radiation moment closures \citep{Wunsch_2024}, and a reduced speed of light approach that becomes increasingly computationally expensive at the high optical depths achieved in this regime of surface densities \citep{Skinner_2013}. Our goal in this paper is to make use of the more accurate radiative transfer methods developed by \citet{Menon_2022} to explore the effects of a top-heavy IMF and varying dust-to-gas ratio in precisely the conditions that JWST is now probing.

In this paper we run idealized radiation hydrodynamic numerical simulations of star cluster formation and their radiative feedback with varying levels of UV luminosity-to-mass (to emulate differing levels of top-heaviness) to quantify how $\epsilon_*$ changes. The paper is organized as follows: In Section~\ref{sec:Methods} we describe the numerical prescriptions we use, and the initial conditions of our clouds and the parameter space we explore.
In Section~\ref{sec:Results} we present the evolution of our model clouds, the $\epsilon_*$ values we find over our parameter space, and discuss the feedback physics driving the trends we find. In Section~\ref{sec:discussion} we provide a discussion on the implications of our results in the context of the JWST results, and enumerate the missing physics in our simulations and their possible effects on our outcomes. In Section~\ref{sec:conclusions} we conclude with a brief summary of our results.

\section{Methods}
\label{sec:Methods}
\subsection{Simulation setup} \label{sec:simsetup}

Our simulation setup is very similar to that described in \citet{Menon_2023}; we briefly summarize the salient features of the setup below and refer the reader that paper for further details. Our simulations represent an isolated cloud of mass $\Mcloud$ and radius $\Rcloud$ which correspond to a mass density of $\rho_{\mathrm{cloud}} = \Mcloud/[(4/3) \pi \Rcloud^3]$ and a mass surface density $\Sigmacloud = \Mcloud/(\pi \Rcloud^2)$. We place our clouds in an ambient medium of density $\rho = \rho_{\mathrm{cloud}}/100$ in pressure-equilibrium in a computational domain of size $L= 4 \Rcloud$. We initialize the fluid with turbulent velocities that follow a power spectrum $E(k) \propto k^{-2}$ with a natural mixture of solenoidal and compressive modes for $k/(2\pi/L) \in \left[2,64 \right]$, generated with the methods described in \citet{Federrath_2010}, and using the implementation of these methods provided in \citet{FederrathEtAl2022ascl}. We scale the velocity dispersion of the cloud $\sigma_v$ such that our clouds are marginally bound, i.e. $\alphavir=2$, where $\alphavir$ is given by
\begin{equation}
    \alphavir = \frac{2E_{\mathrm{kin}}}{E_{\mathrm{grav}}} = \frac{5\Rcloud \sigma_v^2}{3G\Mcloud}\, ,
\end{equation}
with $E_{\mathrm{kin}} = (1/2) \Mcloud \sigma_v^2$ and $E_{\mathrm{grav}} = (3/5) G \Mcloud^2/\Rcloud$. We use diode boundary conditions for the gas quantities wherein we permit gas to escape the boundaries but no inflows through them. 

We model radiation feedback in two wavelength bands, the ultraviolet (UV) and the infrared (IR); the former is technically a combination of the Lyman continuum ($h\nu \geq 13.6 \, \mathrm{eV}$) and Far-UV ($6.6 \leq h\nu < 13.6 \, \mathrm{eV}$) bands, which we collectively refer to as ``UV'' for simplicity. The only sources of UV radiation are  the sink particles that form in our simulations, which represent stellar populations. We adopt a constant UV luminosity-to-mass ratio ($\Psi_{\mathrm{UV}}$) for a given simulation such that the radiative output from a sink of mass $M_{\mathrm{sink}}$ is $L_{\mathrm{UV}} = M_{\mathrm{sink}} \Psi_{\mathrm{UV}}$. On the other hand, the IR emission can come from dust grains that are heated due to the absorption of these UV photons. In addition, we also account for the dust-reprocessed IR field and the associated heating of grains and radiation pressure. We assume that the dust temperature ($T_{\mathrm{d}}$) is instantaneously equal to the radiation temperature set by the equilibrium between dust emission and UV + IR photon absorption \citep[see][for a justification of this assumption]{Menon_2023}. This assumption might cease to hold true in optically thin conditions; in Section~\ref{sec:caveats} we discuss this caveat and argue that it shouldn't affect our results. We use $T_{\mathrm{d}}$ to estimate the Planck/Rosseland emission and absorption opacities in the IR using the temperature-dependent \citet{Semenov_2003} model; \citet{Menon_2022b} show that ignoring this temperature-dependence can strongly overestimate the effectiveness of the IR radiation pressure. For the UV, we assume a fixed opacity (identical Planck and Rosseland opacities) of $\kappa_{\mathrm{UV}} = 1000 \, \mathrm{cm}^2 \, \mathrm{g}^{-1}$ for all our runs\footnote{We did consider the possibility that $\kappa_{\mathrm{UV}}$ for grains exposed to a stellar population with a top-heavy IMF might be different. However, we found that that this effect is very minor as a top-heavy IMF primarily changes the normalisation of the spectrum, and not its shape. This implies luminosity-weighted frequency averages (such as $\kappa_{\mathrm{UV}}$) are very mildly affected. Moreover, as long as our clouds are optically thick in the UV, the radiation force is largely insensitive to the specific value of the UV opacity.}, consistent with typical estimates of the gray radiation pressure cross section per H atom to blackbody radiation peaking at UV wavelengths \citep[blackbody temperatures $\sim$ few $\times 10^4 \, \mathrm{K}$; ][]{Draine_2011,Kim_2023}. These opacities are for $Z = \Zsun$; for other metallicities, we scale our opacities linearly with $Z$ with the underlying assumption that the dust-to-gas ratio scales with metallicity in a linear fashion, which is consistent with observations to zeroth order \citep[e.g.,][]{deVis_2019}. It is possible that this assumption overestimates the dust-to-gas ratio at low $Z$ due to the lack of efficient gas-phase accretion \citep[see, for e.g.,][]{Feldmann_2015,Choban_2022}; however, it shall become clear that a more accurate treatment of the metallicity dependence of the dust-to-gas ratio would only reinforce the conclusions we reach below. We initialize our clouds with zero radiation energy/flux in the UV and an IR radiation field corresponding to a dust temperature of $T_{\mathrm{d}} = 40 \, \mathrm{K}$; this is consistent with dust temperatures in observed high-z starburst galaxies \citep{Sommovigo_2022}. We adopt Marshak boundary conditions \citep{Marshak_1958} for the radiation with the background value set to match the initial conditions. 

We note that we do not include photoionization, stellar winds, protostellar outflows and magnetic fields in these simulations. In Section~\ref{sec:caveats} we discuss (and show in Appendix~\ref{sec:appendix_photoionization}) that the former omission would not affect our results, and discuss the implications of the other missing physics.

\subsection{Parameter space}

All our clouds have $\Mcloud = 10^6 \, \Msun$ with $\Rcloud = 10$ or $3.2 \, \mathrm{pc}$ to achieve a target $\Sigmacloud = 3.2 \times 10^3 \, \Msolpc$ and $3.2 \times 10^4 \, \Msolpc$ respectively. We adopt these values to mimic the high ISM pressure conditions ($P/k_{\mathrm{B}} \sim G\Sigmacloud^2 \sim 10^8 (\Sigmacloud/10^3 \, \Msolpc)^2$) expected, and now being observed at high redshifts (see Section~\ref{sec:intro}). The lower (higher) $\Sigmacloud$ value is (approximately) equal (above) the critical surface density beyond which early stellar feedback is unable to regulate the SFE \citep{Fall_2010,Grudic_2018,Lancaster_2021c,Menon_2023}, which is a key input in models predicting efficient star formation in galaxies such as the FFB model \citep{Dekel_2023}. 

For these two $\Sigmacloud$ cases, we explore variations in $Z$ and $\Psi_{\mathrm{UV}}$ respectively. To mimic increasingly top-heavy IMFs, we explore values of $\Psi_{\mathrm{UV}} = 1,4, \& 10$ times the  value for a standard Chabrier IMF \citep{Chabrier2005} -- i.e. ($\Psi_{\mathrm{Fiducial}} = 900 \, \mathrm{L}_{\sun}\Msun^{-1}$). Parameterizing the top-heaviness of the IMF with $\Psi_{\mathrm{UV}}$ allows us to be agnostic to the degenerate ways in which one can achieve an IMF with an excess of massive stars; regardless, in Appendix~\ref{sec:appendix_slug} we present outputs from the \texttt{SLUG} stellar population synthesis code for how these values map to the slope of the high-mass end of the IMF and/or the maximum mass of the star in the stellar population, for the sake of providing intuition. We note that the $\Psi_{\mathrm{UV}} = 4$ case is additionally motivated by empirical estimates of the factor by which the UV luminosities need to be enhanced to reasonably reproduce the bright end of the UV luminosity functions at $z\sim 10$ with JWST \citep{Yung_2023,Finkelstein_2023}. 
We run each of these cases at $Z = 10^{-2} \Zsun$ in addition to solar metallicity to test the effects of the lower metallicities (and implied lower dust-to-gas ratios) expected at high redshifts. We also run a case with $Z=4 \Zsun$ and $\Psi_{\mathrm{UV}} = \Psi_{\mathrm{Fiducial}}$ motivated by possible evidence of super-solar metallicities found in super-star clusters in our local Universe \citep{Turner_2015}. We summarize our suite of simulations and their parameters in Table~\ref{tab:Simulations}.

\begin{table*}
\caption{Summary of our simulation suite and their initial condition parameters.}
\centering
\label{tab:Simulations}
\begin{threeparttable}
\begin{tabular}{l c c c c c c c c c}
\toprule
\multicolumn{1}{c}{$\Sigma_{\mathrm{cloud}}$}& \multicolumn{1}{c}{$M_{\mathrm{cloud}}$}& \multicolumn{1}{c}{$R_{\mathrm{cloud}}$}&  \multicolumn{1}{c}{$n_{\mathrm{cloud}}$}& \multicolumn{1}{c}{$\sigma_{v}$}& \multicolumn{1}{c}{$v_{\mathrm{esc}}$}& \multicolumn{1}{c}{$t_{\mathrm{ff}}$}& \multicolumn{1}{c}{$\Psi_{\mathrm{UV}}$}& \multicolumn{1}{c}{$\mathrm{Z}$}\\
$[\Msun \, \mathrm{pc}^{-2}]$& $[\Msun]$& [pc]& [$\mathrm{cm}^{-3}$]& [km/s]& [km/s]& [Myr]& [$\Psi_{\mathrm{Fiducial}}$]& [$\Zsun$]&\\
\midrule
\multirow{7}{*}{$3.2$$ \times 10^{3}$} &$10^6$ &$10.0$ &$9.7$$ \times 10^{3}$ &$22$ &29 &0.5 &1 &1\\
&$10^6$ &$10.0$ &$9.7$$ \times 10^{3}$ &$22$ &29 &0.5 &1 &1\\
&$10^6$ &$10.0$ &$9.7$$ \times 10^{3}$ &$22$ &29 &0.5 &4 &1\\
&$10^6$ &$10.0$ &$9.7$$ \times 10^{3}$ &$22$ &29 &0.5 &10 &1\\
&$10^6$ &$10.0$ &$9.7$$ \times 10^{3}$ &$22$ &29 &0.5 &1 &$10^{-2}$\\
&$10^6$ &$10.0$ &$9.7$$ \times 10^{3}$ &$22$ &29 &0.5 &4 &$10^{-2}$\\
&$10^6$ &$10.0$ &$9.7$$ \times 10^{3}$ &$22$ &29 &0.5 &10 &$10^{-2}$\\
&$10^6$ &$10.0$ &$9.7$$ \times 10^{3}$ &$22$ &29 &0.5 &1 &4\\
\midrule
\multirow{7}{*}{$3.2$$ \times 10^{4}$} &$10^6$ &$3.2$ &$3.1$$ \times 10^{5}$ &$40$ &52 &0.09 &1 &1\\
&$10^6$ &$3.2$ &$3.1$$ \times 10^{5}$ &$40$ &52 &0.09 &4 &1\\
&$10^6$ &$3.2$ &$3.1$$ \times 10^{5}$ &$40$ &52 &0.09 &10 &1\\
&$10^6$ &$3.2$ &$3.1$$ \times 10^{5}$ &$40$ &52 &0.09 &1 &$10^{-2}$\\
&$10^6$ &$3.2$ &$3.1$$ \times 10^{5}$ &$40$ &52 &0.09 &4 &$10^{-2}$\\
&$10^6$ &$3.2$ &$3.1$$ \times 10^{5}$ &$40$ &52 &0.09 &10 &$10^{-2}$\\
&$10^6$ &$3.2$ &$3.1$$ \times 10^{5}$ &$40$ &52 &0.09 &1 &4\\


\bottomrule
\end{tabular}
\begin{tablenotes}
\small
\item \textbf{Notes}: Columns in order indicate - Model: $\Sigma_{\mathrm{cloud}}$: mass surface density of the cloud given by $\Sigma_{\mathrm{cloud}} = M_{\mathrm{cloud}}/(\pi R_{\mathrm{cloud}}^2)$, $M_{\mathrm{cloud}}$: mass of cloud, $R_{\mathrm{cloud}}$: radius of cloud, $n_{\mathrm{cloud}}$: number density of the cloud given by $n_{\mathrm{cloud}} = 3M_{\mathrm{cloud}}/(4 \pi R_{\mathrm{cloud}}^3m_{\mathrm{H}})$ where $m_{\mathrm{H}}$ is the mass of atomic hydrogen, $\sigma_{v}$: turbulent velocity dispersion of the cloud, $v_{\mathrm{esc}}$: escape velocity of the cloud, $t_{\mathrm{ff}}$: free-fall time of the cloud, $\Psi_{\mathrm{UV}}$: the UV luminosity per-unit mass scaled by the value for a Chabrier IMF ($\Psi_{\mathrm{Fiducial}}$); see Figure~\ref{fig:luminosity_alpha} for the mapping between the degree of top-heaviness of the IMF and this quantity, $Z$: metallicity in units of solar metallicity ($Z_{\odot}$) -- this is also used to scale the dust abundance assuming a linear trend with $Z$. All our simulations use a resolution of $256^3$. 
\end{tablenotes}
\end{threeparttable}
\end{table*}

\subsection{Numerical Methods}
We solve the equations of self-gravitating radiation hydrodynamics for all our simulations. We use the \texttt{FLASH} magneto-hydrodynamics code \citep{Fryxell_2000,Dubey_2008} for our simulations, with the explicit Godunov method in the split, five-wave HLL5R (approximate) Riemann solver \citep{Waagan_2011} for the hydrodynamics. The Poisson equation for the self-gravity is solved using a multi-grid algorithm implemented in \texttt{FLASH} \citep{Ricker_2008}. Sink particles are used to follow the evolution of gas at unresolved scales, the formation of which is triggered when gas properties satisfy a series of conditions to test for collapse and star formation \citep{Federrath_2010_Sinks}. Gravitational interactions of sink particles with gas and other sinks are considered, and a second-order leapfrog integrator is used to advance the sink particles \citep{Federrath_2010_Sinks,FederrathBanerjeeSeifriedClarkKlessen2011}. To model the radiative transfer and the associated energy and momentum transfer to gas, we use the Variable Eddington Tensor-closed Transport on Adaptive Meshes method (\texttt{VETTAM}; \citealt{Menon_2022}). \texttt{VETTAM} solves the non-relativistic, angle-averaged, moment equations of radiative transfer in the mixed-frame formulation \citep{Mihalas_1982}, retaining terms that are of leading order in all limiting regimes of RHD \citep[see, e.g.,][]{Krumholz_2007a}. It uses the VET closure obtained with a time-independent ray-trace solution \citep{Buntemeyer_2016} to close the moment equations; this approach yields much more accurate solutions for problems with multiple radiation sources than any purely local approximation for the Eddington tensor (e.g., the M1 approximation). \texttt{VETTAM} uses an implicit global temporal update for the radiation moment equations for each band, accounting for the coupling between the bands due to dust-reprocessing. The radiative output from sink particles is included as a smoothed source term in the moment equations, where we have tested convergence in the smoothing parameters \citep[see][]{Menon_2022b}. We used a fixed uniform grid resolution of $256^3$ for all our simulations; although \texttt{VETTAM} fully supports AMR, we chose a fixed modest resolution for simplicity and the computational feasibility required to explore our broad parameter space; we demonstrate convergence in the SFE (within $\lesssim 5 \%$) at these resolutions in similar numerical setups in \citet{Menon_2023}.

We pause to comment that our numerical model has been used to study the competition between star formation and feedback set by radiation pressure in our previous work \citep{Menon_2022b,Menon_2023}. These works showed that the integrated star formation efficiency increases with the gas surface density of the cloud, producing efficiencies approaching unity for $\Sigmacloud \gtrsim 10^4 \, \Msolpc$. However, for gas surface densities representative of the local Universe ($\Sigmacloud \sim 100 \Msolpc$), we found efficiencies $\sim 30 \%$, which is substantially higher than other works in the literature \citep[e.g.,][]{Kim_2017}. We can confirm that this was simply because this work did not include photoionization, which becomes important in that parameter regime; indeed in our more recent work with photoionization (Menon et al., in prep) we find efficiencies $\sim 10\%$ for clouds in this parameter regime -- in strong agreement with other numerical simulations and observed estimates \citep{Chevance_23}. We make this point here to clarify for the reader that our numerical model produces consistent results in regions of parameter space that have been studied widely in previous work.

\section{Results}
\label{sec:Results}

\subsection{Competition between star formation and feedback}

\begin{figure*}
    \centering
    \includegraphics[width=\textwidth]{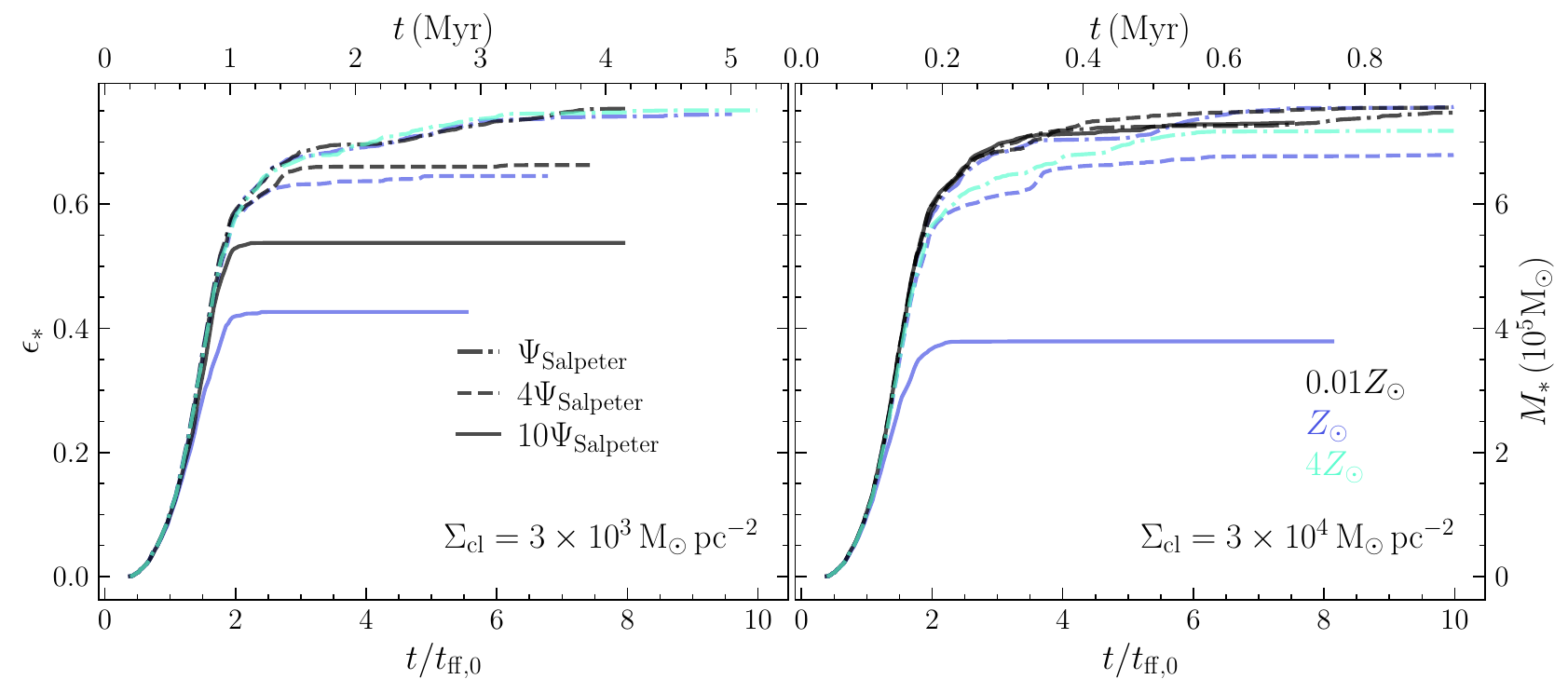}
    \caption{The star formation efficiency as a function of time for our simulations; $\Psi_{\mathrm{UV}} = 1,4, \& 10 \, \Psi_{\mathrm{Fiducial}}$ are represented by dashdot, dashed, and solid lines respectively, colours indicate the different metallicities we explore, and the two panels represent the different $\Sigmacloud$ values in our simulation suite (Table~\ref{tab:Simulations}). All our simulations undergo rapid collapse (over $t \lesssim 2 t_{\mathrm{ff}}$) and star formation followed by a saturation in $\epsilon_*$ at the point when feedback is able to counteract the collapse; this saturation point is clearly different across our runs.}
    \label{fig:SFEvsTime}
\end{figure*}

\begin{figure*}
    \centering
    \includegraphics[width=\textwidth]{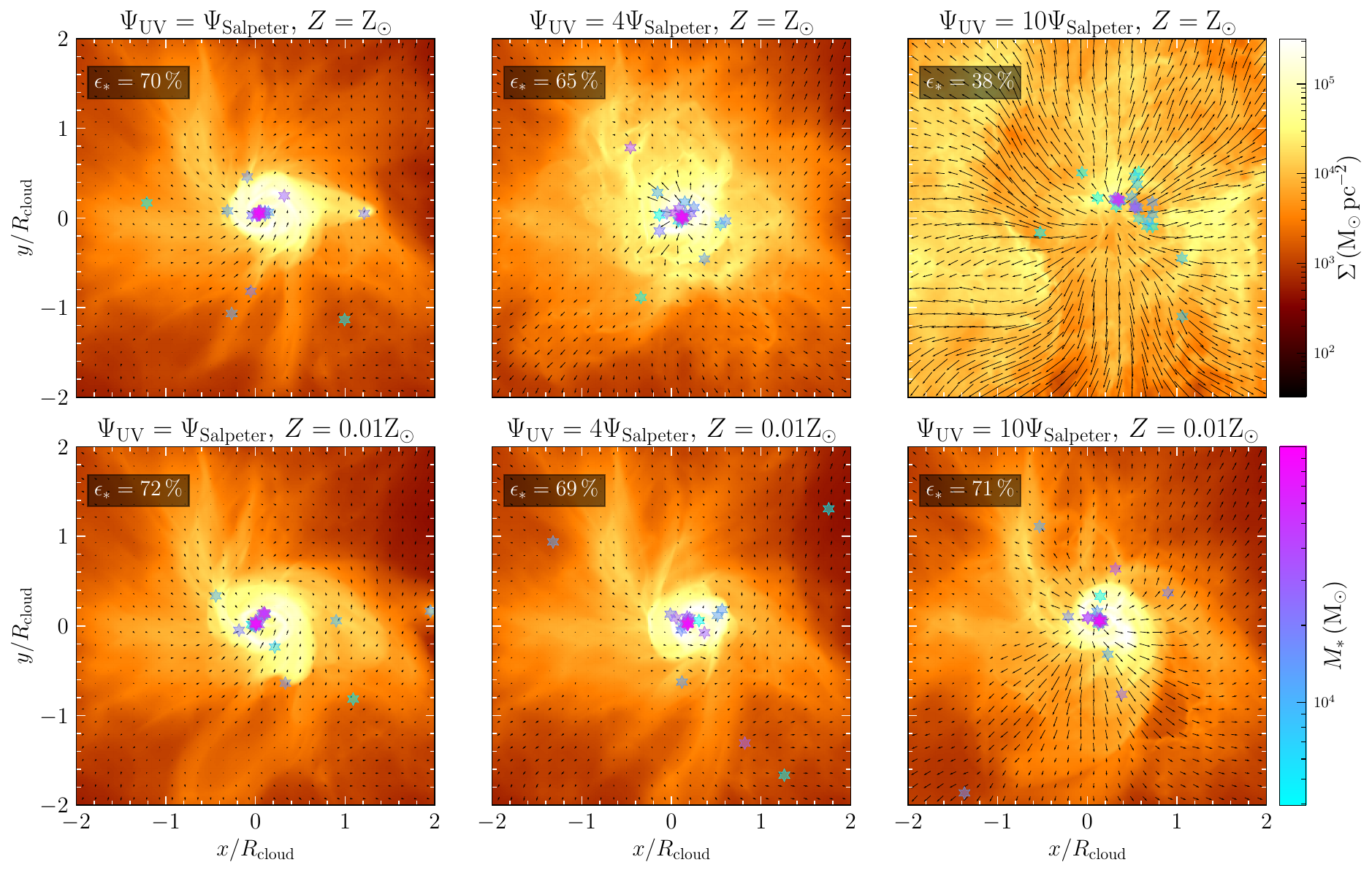}
    \caption{Gas surface density distributions at a time $t=4 t_{\mathrm{ff}}$ for the $\Sigmacloud = 3.2 \times 10^4 \, \Msolpc$ runs with increasing $\Psi_{\mathrm{UV}}$ (left to right) at $Z = Z_{\odot}$ (top) and $Z=0.01 Z_{\odot}$ (bottom). We can see the general trend of a stronger impact of feedback at higher $\Psi_{\mathrm{UV}}$ and $Z$, as evidenced by presence/absence of outflows in the velocity distribution, which are driven by radiation pressure on dust. The star formation efficiency ($\epsilon_*$; annotated top-left of each panel) is high in all these cases, except when both $\Psi_{\mathrm{UV}}$ and $Z$ are high. Comparison of the achieved $\epsilon_*$ in the top-right and bottom-right panels demonstrate that a drop in $Z$ can counteract the effects of a higher $\Psi_{\mathrm{UV}}$.}
    \label{fig:GasProjection}
\end{figure*}

The initial evolution of all our models are relatively similar -- density enhancements due to the turbulent fluctuations undergo self-gravitational collapse and go on to form sink particles (that represent sub-clusters), which then accrete and continue to increase the total stellar mass (and therefore $\epsilon_* = M_*/\Mcloud$) in the cloud. Turbulent fluctuations also introduce some non-negligible mass-loss through the computational boundaries at early times ($\sim 10-20 \%$). This occurs because local gas patches can become unbound and escape through our isolated boundary conditions in the initial phases, although the cloud is globally marginally stable. The stellar mass continues to grow rapidly for $t\lesssim t_{\mathrm{ff}}$ after which the evolution between the clouds start to differ due to the regulating effects of radiative feedback. We can see this in Figure~\ref{fig:SFEvsTime} which shows the time evolution of $\epsilon_*$ for all our model clouds. The rate of star formation -- interpreted from the rate of change of $\epsilon_*$ with time -- for clouds with progressively higher $\Psi_{\mathrm{UV}}$ slows down earlier and more dramatically. This is due to the stronger feedback around the radiating sources that reverses the accretion flow in its vicinity, and starts to drive this gas locally outward. We can see this visually in Figure~\ref{fig:GasProjection} which show the projected gas density and velocity fields at $t = 3 t_{\mathrm{ff}}$ for the higher $\Sigmacloud$ runs. We can see that the sinks are still accreting for $\Psi_{\mathrm{UV}} = \Psi_{\mathrm{Fiducial}}$, whereas increasing amounts of gas are outflowing for higher $\Psi_{\mathrm{UV}}$, and the resulting $\epsilon_*$ is lower. This can be understood due to the stronger levels of UV radiative feedback for these cases at any given stellar mass. However, consistent with Figure~\ref{fig:SFEvsTime}, the $Z \sim 0.01 \Zsun$ cases seem to show much more modest effects from the feedback even for the higher $\Psi_{\mathrm{UV}}$ cases. The stellar masses accumulated in the same time are also higher. This suggests that the lower dust-to-gas ratio in these runs skews the feedback-star formation competition in favour of the latter. We can see that the effects of the dust-to-gas ratio are much less pronounced in the lower $\Sigmacloud$ case. We explain the reason for this behavior with $Z$ and $\Sigmacloud$ in Section~\ref{sec:physics_trends}. 

\subsection{Integrated Star formation efficiencies}

\begin{figure*}
    \centering
    \includegraphics[width=\textwidth]{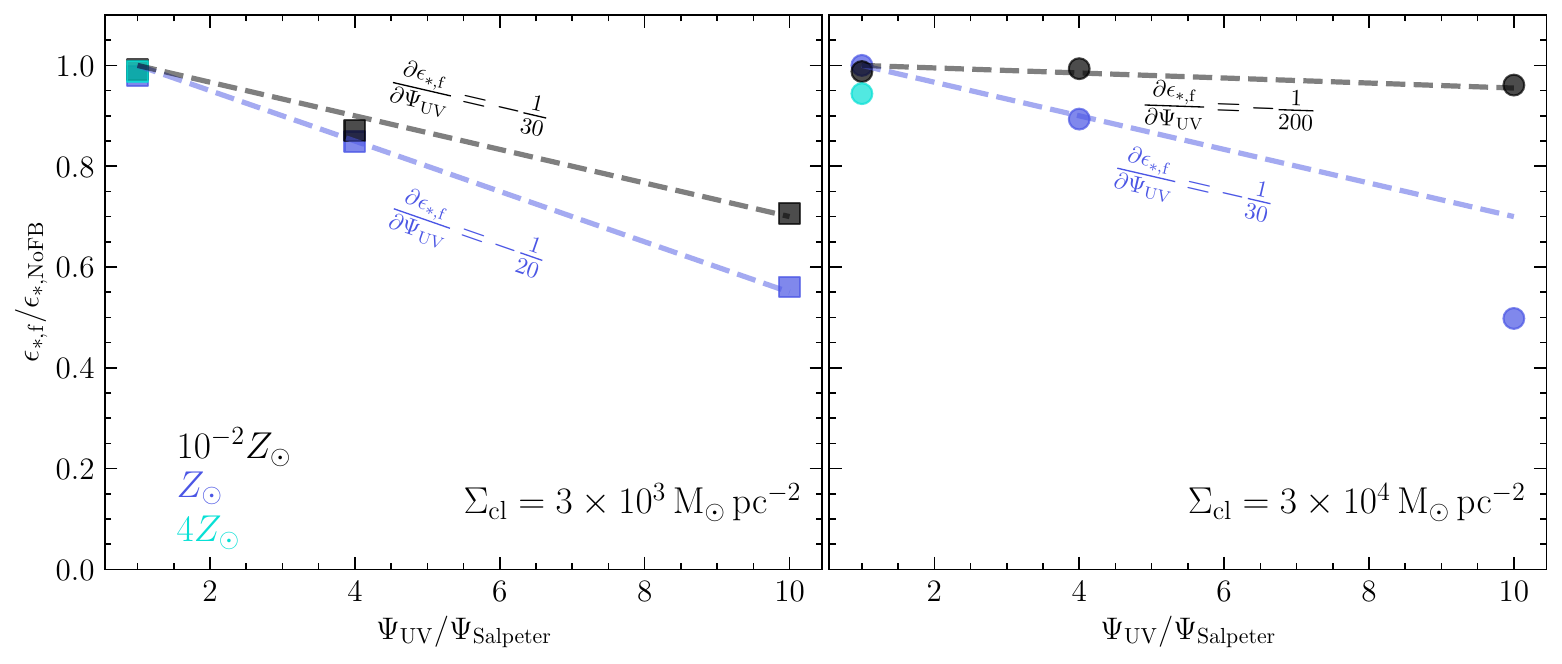}
    \caption{The final integrated cloud scale star formation efficiency ($\epsf$) obtained in all of our simulations -- scaled by the value obtained for a run without feedback ($\epsilon_{*,\mathrm{NoFB}}$) -- shown as a function of $\Psi_{\mathrm{UV}}$. We indicate lines and their slopes to guide the eye (no fitting). We can see that there is a general trend of decreasing $\epsf$ with $\Psi_{\mathrm{UV}}$, however, a lower $Z$ can at least partly compensate for this decrease -- much more so for the higher $\Sigmacloud$ cases (right panel). The trend for the higher $\Sigmacloud$ case is also clearly non-linear. These trends are likely driven by the differing levels of momentum imparted by radiation pressure on dust across our simulations (see Section~\ref{sec:physics_trends}).}
    \label{fig:finalSFE}
\end{figure*}

The aforementioned trends are also reflected in the final saturated level of $\epsilon_{*}$ set by the star formation/feedback balance in our simulations -- our key quantity of interest. We calculate this as the value at the point when there is less than $5 \%$ of the gas mass remaining in the computational domain, and is indicated by the termination of the curves in Figure~\ref{fig:SFEvsTime}. When reporting this quantity we normalize by its corresponding value for a control run without any feedback to account for the initial gas mass loss due to our isolated turbulent cloud numerical setup, since this gas does not participate in the feedback-star formation competition; we refer to this normalised final star formation efficiency as $\epsf$. We also do this to place less emphasis on the \textit{exact} value of $\epsilon_*$ in the simulation -- since this is expected to vary depending on the turbulent initial conditions -- but rather on the relative effect of the feedback for a given cloud. We show the values obtained for $\epsf$ across our simulation suite in Figure~\ref{fig:finalSFE} as a function of the input value of $\Psi_{\mathrm{UV}}$ we use for the stellar populations. As expected, we can see the general trend that $\epsf$ decreases with increasing $\Psi_{\mathrm{UV}}$. However, $\epsf$ increases with decreasing $Z$ for a given $\Psi_{\mathrm{UV}}$. The dependence on $Z$ is weak for $\Sigmacloud \sim 10^3 \, \Msolpc$ ($\lesssim 10 \%$). However for the higher $\Sigmacloud$ case, the lower dust content can more or less completely counteract the effects of the higher $\Psi_{\mathrm{UV}}$. We can also see that $\epsf$ for the run with $Z \sim 4 \Zsun$ is almost identical to the solar metallicity run; for the higher $\Sigmacloud$ case the difference is slightly more evident. This, along with the corresponding trends for lower metallicities, suggests that the dependence on dust content is stronger for the higher $\Sigmacloud$ case.

We also overplot approximate trends with $\Psi_{\mathrm{UV}}$ to guide the eye for $Z=\Zsun$ and $Z=0.01 \Zsun$. We can see that the trend is largely linear (albeit with different slopes) for the lower $\Sigmacloud$ run, whereas it is clearly non-linear for the higher $\Sigmacloud$ at $Z=\Zsun$, but essentially flat for $Z=0.01 \Zsun$. This implies that for this case, the lower dust-to-gas ratio completely compensates for the (high) increase 
in the UV luminosity. This has important implications for star cluster formation at high redshifts, as we will discuss below.

\subsection{Physics driving trends}
\label{sec:physics_trends}

In this section we briefly explain the feedback physics that drives the trends with $\Psi_{\mathrm{UV}}$ and $Z$ in our simulations. The primary feedback mechanisms that drives the dynamics in our clouds is the radiation pressure on dust grains -- both the single-scattering UV force, and the multiple-scattering force due to re-emitted IR radiation by warm dust. The former applies a constant force over the absorbing shell of $\sim L_*/c$ as long as it is optically thick in the UV; this requires $\Sigma \gtrsim 5 (Z/\Zsun)^{-1} \Msolpc$, which is satisfied across our parameter space even for the $Z= 0.01 \Zsun$ runs. On the other hand, to be optically thick in the IR requires $\Sigma \gtrsim 10^3 (Z/\Zsun)^{-1} \Msolpc$ assuming an average $\kappa_{\mathrm{IR}} = 5 \, \mathrm{cm}^2 \, \mathrm{g}^{-1}$. Of the simulations we present in this paper, only the two with $\Sigmacloud \gtrsim 10^4 \, \Msolpc$ with $Z = \Zsun$ and $4\Zsun$ satisfy this condition. These points imply that the IR radiation pressure is only an important contributor for these subset of runs -- this is consistent with the findings reported in \citet{Menon_2023}. In these conditions, the trapped IR radiation field can impart a force $\sim \ftrap L_*/c$, where $\ftrap$ describes the trapping factor that quantifies the momentum gained by the multiple scattering of the IR radiation.

Now the competition between radiation pressure and gravity can be quantified with the Eddington ratio. The Eddington ratio for a column of gas with surface density $\Sigma$ exposed to a stellar population with UV luminosity-per unit mass $\Psi_{\mathrm{UV}}$ for single-scattering radiation pressure is \citep[e.g.,][]{Thompson_2015}

\begin{equation}
    \label{eq:GammaUV}
    \Gamma_{\mathrm{UV}} = \frac{\Psi_{\mathrm{UV}}}{4 \pi G c \Sigma} \sim  0.1 \left(\frac{\Psi_{\mathrm{UV}}}{\Psi_{\mathrm{Fiducial}}} \right) \left(\frac{\Sigma}{3000 \, \mathrm{M}_{\odot}\, \mathrm{pc}^{-2}} \right)^{-1}.
\end{equation}
Since there would be a distribution of $\Sigma$ surrounding the stellar population as the cloud evolves \citep{Thompson_Krumholz_2016}, the above calculation suggests that the fraction of sightlines that become super-Eddington increases with $\Psi_{\mathrm{UV}}$. This could, to zeroth order, explain the trend we see in the $\Sigmacloud = 3000 \, \Msolpc$ runs. Note that there is no dependence on $Z$ for $\Gamma_{\mathrm{UV}}$, as long as the gas is optically thick in the UV.

On the other hand, for the multiple-scattering IR radiation force, the Eddington ratio is 
\begin{equation}
    \label{eq:GammaIR}
    \Gamma_{\mathrm{IR}} = \frac{\kappa_{\mathrm{IR}}\Psi_{\mathrm{UV}}}{4 \pi Gc} = 0.4 \left(\frac{\kappa_{\mathrm{IR}}}{5 \, \mathrm{cm}^2 \, \mathrm{g}^{-1}}\right)\left(\frac{\Psi_{\mathrm{UV}}}{\Psi_{\mathrm{Fiducial}}} \right) \left(\frac{Z}{\Zsun} \right),
\end{equation}
where we explicitly note the (assumed) linear dependence of $\kappa_{\mathrm{IR}}$ -- the IR opacity -- on the metallicity\footnote{We stress that this expression is valid only if the gas is optically thick in the UV such that the full stellar luminosity gets reprocessed to the IR. In addition, this expression amounts to assuming $\ftrap = \kappa_{\mathrm{IR}} \Sigma$ for the IR radiation force; in reality, $\ftrap$ would depend on the dust temperatures through the column of gas \citep{Menon_2022b} and nonlinear radiation-matter interactions \citep{Krumholz_Thompson_2012} -- both of which is captured in our simulations. The constant $\kappa_{\mathrm{IR}}$ value we use in this expression is just a simplification we make to explain the qualitative trends we find in our simulations.}. Note that this does not have a dependence on $\Sigma$, as long as the column is optically thick in the IR, i.e. $\Sigma \gtrsim 10^3 (Z/\Zsun)^{-1} \Msolpc$. 

We can now understand the stronger metallicity dependence for the higher $\Sigmacloud$ case: it is in the regime where the dust-to-gas ratio dependent IR radiation pressure is the dominant feedback mechanism. This force only plays a relatively minor role\footnote{However, this effect is not negligible --- some sightlines can become optically thick in the IR due to the turbulent overdensities. This likely explains the $\sim 10 \%$ differences between the $Z=\Zsun$ and $Z=0.01 \Zsun$ runs for this cloud.} for $\Sigmacloud \lesssim 10^4 \, \Msolpc$ -- even less so at lower $Z$ -- as it is optically thin in the IR. For the higher $\Sigmacloud$ case the IR radiation pressure is clearly the crucial force as the UV radiation pressure has insufficient momentum to compete with gravity at this high $\Sigma$ -- even for $\Psi_{\mathrm{UV}} = 10 \Psi_{\mathrm{Fiducial}}$. This is reflected in the value of $\epsf$ for $Z=0.01 \Zsun$ where only the UV acts; an inspection of Equation~\ref{eq:GammaUV} clearly indicates sub-Eddington conditions for these parameters. The linear dependence on $\Psi_{\mathrm{UV}}$ in Equation~\ref{eq:GammaUV} also explains the linear trend seen with $\Psi_{\mathrm{UV}}$ for the lower $\Sigmacloud$ cases. Finally, we find that the non-linear trend seen at $Z=\Zsun$ for the higher $\Sigmacloud$ runs is due to the subtle effect of more efficient trapping of IR photons -- i.e. higher $\ftrap$ -- for clouds with higher $\Psi_{\mathrm{UV}}$. This occurs because the (IR) radiation temperature is significantly higher for higher $\Psi_{\mathrm{UV}}$, which renders $\kappa_{\mathrm{IR}}$ higher, thereby imparting more momentum per unit stellar mass. Connecting with Equation~\ref{eq:GammaIR}, it is the combination of increasing $\kappa_{\mathrm{IR}}$ due to warmer dust along with the linear increase due to $\Psi_{\mathrm{UV}}$ that leads to the non-linear trend.

\section{Discussion}
\label{sec:discussion}

\subsection{Implications for massive galaxies at cosmic noon}
Higher global star formation efficiencies than the local Universe and a top-heavy IMF are two of several proposed scenarios to reconcile the observed abundance of massive UV-bright galaxies at $z \sim 8$--$12$  with pre-launch model predictions \citep{Inayoshi_2022,Harikane_2023,Finkelstein_2023,Yung_2023}. The former is a key element of the Feedback-Free Burst (FFB) model \citep{Dekel_2023}, which invokes high cloud-scale SFE and ineffective feedback by stellar and supernova driven winds to achieve more efficient galaxy-scale star formation and hence boost the numbers of UV-luminous galaxies at early times. 
\citet{Li_2023} showed that this model is consistent with observations when they adopt cloud-scale SFE of $\lesssim 50\%$. On the other hand, a top-heavy IMF and the associated higher UV luminosities could also match the UV luminosity functions while still adopting the lower SFEs that seem to be typical of galaxy populations at $z \lesssim 8$ \citep{Tacchella_2018}; \citet{Yung_2023} show that a boost of $\sim 4$ in the UV luminosity-to-mass ratio can reproduce the UV luminosity function at $z\sim 11$ without modifying the star formation efficiency or feedback strength. However, these studies explored the impact of the SFE and the IMF as if they were independent, which of course is not the case in reality. In this study, we have quantified how these two quantities depend on each other at the cloud scale.

Our results indicate that as long as clouds have surface densities $\Sigmacloud \gtrsim 10^3 \, \Msolpc$ -- a condition that seems to be commonly satisfied at $z\gtrsim 10$ based on observed galaxy sizes \citep[e.g.,][]{Finkelstein_2023,Adamo_2024,Casey_2023,Morishita_2023} -- a star formation efficiency significantly higher than that typical of the Local Universe ($\sim 10\%$) is unavoidable \emph{even in the presence of a top-heavy IMF} (we have investigated cases where the luminosity-to-mass ratio is up to ten times the typical value; see Figure~\ref{fig:luminosity_alpha} and Section~\ref{sec:appendix_slug}). A top-heavy IMF results in only a moderate reduction in the star formation efficiency, and only if the dust abundance is similar to the solar neighborhood. For metallicites that seem to be typical at the highest redshifts where we have reliable estimates, $z \sim 8$--10, i.e. $Z \sim 0.1$-0.3 $\Zsun$
\citep{Curti_2023,Nakajima_2023},
and assuming a linear relation between dust-to-gas ratio and metallicity\footnote{Dust-to-gas ratios are highly uncertain at the masses and redshifts of the $z\gtrsim10$ galaxies. Observational constraints on dust-to-gas ratios at $z\sim 2$ and metallicities of  12+$\log$(O/H) $\sim 8.5$--8.8 are consistent with values in the nearby Universe. However, 
for local galaxies, \citet{remy-ruyer:2014} find that the relationship between dust-to-gas and metallicity is best fit by a broken double power-law, while \citet{deVis_2019} find it is well fit by a single power law. This could lead to a difference of up to an order of magnitude in dust-to-gas at the typical metallicities (12+$\log$(O/H) $\sim 7.5$) of the JWST galaxies. }, the SFE is higher for a given $\Psi_{\mathrm{UV}}$, and completely counteracts the effects of $\Psi_{\mathrm{UV}}$ for highly compact clouds ($\gtrsim 10^4 \, \Msolpc$). Moreover, the nature of the IMF for conditions at $z\sim 10$ is highly uncertain, with works suggesting that it could even be bottom-heavy \citep[e.g.,][]{Conroy_2012,Tanvir_2022,Tanvir_2024}; that being said, these studies probe the IMF in a mass range ($\lesssim 1 \Msun$) that does not contribute to $\Psi_{\mathrm{UV}}$. Even if this scenario were true for the high-mass end of the IMF, it would only imply that even more efficient star formation would be required, as $\Psi_{\mathrm{UV}}$ would be even lower than with a standard IMF. All of this suggests that highly efficient star formation at the cloud scales may be ubiquitous in high redshift galaxies, irrespective of the properties of the stellar populations that populate them. 

If we take this at face value, it is possible that the combination of a top-heavy IMF in addition to efficient star formation could \textit{overpredict} the UV luminosity functions, since they both contribute to an excess at the bright end. 
However, there are two key subtleties to point out in this context. Firstly, studies that found that a factor $\sim 4$ increase in $\Psi_{\mathrm{UV}}$ is sufficient to reproduce the observations do not account for any possible dust extinction \citep[e.g.,][]{Yung_2023}. Secondly, the SFE values and trends we quantify in this study are at the cloud scale ($\lesssim 10 \, \mathrm{pc}$), whereas the quantity relevant for the luminosity functions is the baryon efficiency ratio defined over the whole galaxy ($\epsilon_{*,\mathrm{gal}}$). For instance, \citet{Li_2023} find that the FFB model fits the tentative JWST data for $\epsilon_{*,\mathrm{gal}} \sim 20\%$. This could either reflect the true SFE values within each star-forming cluster or alternatively reflect a duty cycle of star formation in the galaxy. The FFB scenario does predict a duty cycle\footnote{We note that the episodic nature of the star-formation history by itself can help bias the luminosities upwards in the bright end where the luminosity function is steep, allowing fits to the observations with even lower $\epsilon_{*,\mathrm{gal}}$ \citep{Sun_2023}.}, due to the need to accumulate enough accreted gas for triggering the fragmentation into star-forming clouds, which can lead to an $\epsilon_{*,\mathrm{gal}} \sim 20\%$ in spite of the assumed SFE $\sim 100 \%$ at the scale of the individual clouds \citep{Dekel_2023,Li_2023}. The results obtained in the current work of higher SFE within the individual clusters is consistent with this duty-cycle interpretation of the lower SFE when averaged over time in the galaxy. 

Alternatively, another possibility to reconcile our high cloud-scale SFE values with relatively lower $\epsilon_{*,\mathrm{gal}}$ is that only a fraction of the gas in the galaxy participates in star formation in clouds. It is possible the remaining gas is ejected in outflows by the feedback from older stellar populations, possibly explaining the dust-free nature of these galaxies \citep{Fiore_2023,Ferrara_2023a,Ferrara_2023b}, which is likely critical to simultaneously explain the observed UV luminosity functions and the blue UV continuum slopes \citep[e.g.,][]{Cullen_2023} -- although see \citet{Li_2023} for an alternative explanation of these findings. This possibility raises another effect of a top-heavy IMF that we cannot capture in our simulations: a top-heavy IMF\footnote{This effect would depend on the shape of the IMF and how top-heaviness is achieved -- for instance, this would not apply if the top-heaviness comes from the upper mass cuttoff of the IMF being higher, since these stars would not undergo supernovae.}, would lead to more energy and mass loaded winds, potentially further decreasing $\epsilon_{*,\mathrm{gal}}$, such that its combination with a top-heavy IMF could be consistent with the observations. There is scope for studying the interaction of these three key parameters -- SFE at the cloud scale, IMF, and the feedback effects in driving galaxy-scale winds, taking into account their respective dependencies on each other. By combining these joint constraints into a galaxy-scale semianalytic model \citep[SAM;][]{Somerville_2015}, we may be able to constrain the regions of parameter space permitted by the observations. 

\subsection{Missing physics and possible implications}
\label{sec:caveats}

It is important to note that several physical mechanisms are missing in our numerical simulations -- we list them here and discuss how they might affect the outcomes. 

We only model the radiative feedback on dust, and do not include photoionization and therefore the momentum from the associated thermal pressure of ionized gas. However, we argue that this would make little difference to the outcome of our simulations, as the clouds we model have escape speeds $v_{\mathrm{esc}} \gtrsim$ \mbox{$\sim 2$--$5c_{s,\mathrm{ion}}$} where $c_{s,\mathrm{ion}} \sim 10 \, \mathrm{km} \, \mathrm{s}^{-1}$ is the ionized gas sound speed\footnote{We also verify that a top-heavy IMF does not lead to higher $c_{s,\mathrm{ion}}$. The reason is identical to footnote 2: the heating rate on H-atoms absorbing UV photons does not change due to a very similar spectrum. That being said, a lower $Z$ could lead to a higher gas temperature by a factor $\sim 2$ due to decreased metal-line cooling in photoionized gas; however this would imply a marginal increase to a value of $c_{s,\mathrm{ion}} \sim 14 \, \kms$.}. These arguments are consistent with results presented in models and previous numerical simulations that show that radiation pressure on dust is the dominant radiative feedback mechanism in this regime for regulating star formation \citep{Krumholz_Matzner_2009,Dale_2012,Kim_2016,Kim_2018}. We also demonstrate this is the case in Appendix~\ref{sec:appendix_photoionization} by re-running one of our models with the effects of photo-ionization included. We can justify the omission of protostellar outflows along similar lines -- $v_{\mathrm{esc}}$ is much higher than the $\sim 1 \kms$ threshold suggested by \citet{Matzner_2015} for effective gas ejection by jets. 

We also do not model stellar winds. While this might at first seem like a major omission, we note that the effectiveness of stellar wind feedback has been shown to be reduced compared to analytic estimates due to efficient cooling at turbulent interfaces in the multiphase gas, rendering it momentum-limited \citep{Lancaster_2021a,Lancaster_2021b}; this has been shown to be especially true in the regime of high $\Sigmacloud \gtrsim 10^3 \, \Msolpc$ \citep{Lancaster_2021c} that we focus on here. However, this still implies that there would be a force $\dot{p}_{\mathrm{w}}$ -- where $\dot{p}_{\mathrm{w}} \sim \dot{M_{\mathrm{w}}} v_{\mathrm{w}}$ is the wind momentum for a mass-loss rate $\dot{M_{\mathrm{w}}}$ and wind-velocity $v_{\mathrm{w}}$ -- acting on the gas. $\dot{p}_{\mathrm{w}}$ is expected to be $\sim L_*/c$ for a stellar population \citep[see Figure 3, ][]{Lancaster_2021a}, suggesting that this should induce an order unity correction to our obtained values of $\epsf$. In other words, it is possible that the $\epsf$ obtained at $\Psi_{\mathrm{UV}} = 10$ would be obtained for  $\Psi_{\mathrm{UV}} = 5$ with the additional effect of stellar winds. That being said, it is highly likely that the two feedback mechanisms do not interact in a simple additive fashion. Moreover, at metallicities $Z \lesssim 0.1 \Zsun$ winds from massive O stars are considerably weaker \citep{Leitherer92,Vink01}, meaning that significant stellar wind feedback is delayed until the onset of Wolf-Rayet winds \citep{Lancaster_2021a,Dekel_2023}. This time delay may be too long to have a significant impact on star formation in clouds of these densities. Numerical simulations that combine wind and radiative feedback would provide more formal quantification of the resultant star formation efficiencies in such conditions. 

We assume perfect coupling between gas and dust temperatures, and radiative equilibrium for the radiation and dust temperatures. These assumptions are quite reasonable when our clouds are optically thick in the IR; however, for our simulations with $Z \sim 0.01 \Zsun$, they start to break down. For instance, dust and gas temperatures likely decouple in these conditions except in very high density regions ($n \gtrsim 10^7 \mathrm{cm}^{-3}$), which renders our estimates for the gas temperatures incorrect. However, the dynamical impact of this would be minor, since the thermal pressure is not a significant force in the systems we are investigating. The fragmentation properties in our simulations would be affected by this error, but we do not resolve individual stars anyway and our scope is limited to studying the net competition between radiation forces and gravity in clouds, which is unlikely to be affected. 

Our assumption that the dust and the radiation field are in LTE also starts to break down at low dust-to-gas ratios when the dust becomes optically thin in the IR. In this limit, the color temperature of the IR radiation field at any spatial location is not equal to the local dust temperature -- an effect that can only be captured by a numerical method that models the evolution of the full SED through the cloud. The way in which this assumption directly affects our numerical model is that our estimated dust temperatures would be incorrect in the optically-thin limit, directly affecting the IR dust absorption opacities, which then subsequently affects the IR component of the radiation force (which is $\propto$ opacity). However, we estimate that the impact of this would be minor, since this error only applies in the limit where the dust is optically thin in the IR, in which case we are in the single scattering regime anyway and the IR radiation force is negligible; the latter becomes important only when optically thick in the IR, in which case our assumption is valid. One might question if the very statement that the cloud is optically thick/thin in the IR might itself be affected by the (indirect) error we make in the dust opacity. We estimate that this is unlikely since the range of (gray) IR opacities for dust warmer than 40 K varies by at most a factor $\sim$ few \citep[see Figure 1;][]{Menon_2022b}. Therefore even if we assume significant error in the dust temperature (which is itself unlikely\footnote{This is again due to the identical reason of the relatively narrow range of the IR dust opacities; even if the color temperature of the IR radiation be significantly differently from the local dust temperature ($T_\mathrm{d}$), the corresponding opacity can at most be different by a factor say $f \sim$ few. Since $T_{\mathrm{d}}/T_{\mathrm{d,LTE}} \propto f^{1/4}$, this would change $T_{\mathrm{d}}$ from the LTE value by a factor $\lesssim 2$.}), it results in the IR dust opacity being underestimated by a factor $\sim$ few, which is insufficient to alter the regime of the problem from the single scattering to the multiple scattering regime for our $Z = 0.01 \Zsun$ clouds, which are optically thin by at least 1-2 orders of magnitude. Hence, the impact of this assumption on our results are unlikely to be significant. That being said, this is a subtle effect that could affect systems that are marginally optically thick; there is scope for future (frequency-dependent) calculations to quantify the impact of this assumption in such conditions.

We also do not include magnetic fields which could provide additional support against gravitational collapse and therefore possibly render higher fractions of gas unbound \citep[e.g.,][]{Burkhart_2018,Krumholz_Federrath_2019,Kim_2021}. In addition, we do not have the influence of an external larger-scale turbulent environment which could provide additional stabilization \citep{Kim_2021,Orr_2022,Forbes_2023} through a turbulent cascade acting on the scales of our clouds but also possibly additional compressive modes \citep{Appel_2023}. Both of these could slightly affect our obtained values of $\epsf$. We therefore urge caution in interpreting the exact values of $\epsf$ we report. We emphasize our main takeaway is the trends we find with the IMF and dust content (metallicity). 

\section{Conclusions}
\label{sec:conclusions}
We study the efficiency of star formation set by radiative feedback for assumed IMFs that are (increasingly) top-heavy and at different dust-to-gas ratios (or metallicity $Z$, assuming a linear relation between the two). We focus on massive, dense, compact clouds with initial gas surface densities $\Sigmacloud \gtrsim 10^3 \, \Msolpc$, which are likely typical for galaxies that have been detected by JWST at $z\sim 10$. Past theoretical studies have shown that clouds in this regime are expected to exhibit very high star formation efficiencies \citep[e.g.,][]{Lancaster_2021c,Menon_2023,Polak_2023} for a standard UV luminosity-to-mass assuming a \citet{Chabrier2005} IMF ($\Psi_{\mathrm{Fiducial}}$). We test the effects of increased feedback due to a top-heavy IMF on such clouds, by assuming different values of the UV luminosity-to-mass ratios ($\Psi_{\mathrm{UV}}$) -- up to $10 \Psi_{\mathrm{Fiducial}}$ -- for the stellar populations forming in our simulations. We also explore the effects of sub- and super-solar metallicities to mimic different environments/redshift conditions (see Table~\ref{tab:Simulations}). Our takeaway findings are:
\begin{enumerate}
    \item \textit{Efficient Star Formation}: The integrated cloud-scale star formation efficiency ($\epsf$) is $\gtrsim 40\%$ in all our simulations in spite of very top-heavy IMFs. This is much higher than that typical of young star clusters in the Local Universe ($\epsf \lesssim 10\%$), suggesting that (relatively) efficient star formation is unavoidable at high gas/stellar surface densities.
    
    \item \textit{Effects of Top-Heavy IMF}: Efficiencies are lower for an increasingly top-heavy IMF  (i.e., higher $\Psi_{\mathrm{UV}}$). We find that for solar-neighborhood dust conditions, $\epsf$ decreases with $\Psi_{\mathrm{UV}}$ by up to $50 \%$ ($20\%$) for an increase in $\Psi_{\mathrm{UV}}$ by a factor 10 (4). 
    
    \item \textit{Effects of Dust-to-Gas Ratio}: Efficiencies are higher for sub-solar metallicity/dust abundance, which seems to at least partly compensate for the effects of a top-heavy IMF (Figure~\ref{fig:finalSFE}). This effect is much stronger for more compact clouds ($\Sigmacloud \gtrsim 10^4 \, \Msolpc$) such that the star formation efficiency is indistinguishable from the fiducial case even at $\Psi_{\mathrm{UV}} = 10 \Psi_{\mathrm{Fiducial}}$. At $\Sigmacloud \sim 10^3 \, \Msolpc$ the effect of the dust content are more modest ($\lesssim 10\%$). 
    
    \item \textit{Radiative Feedback}: We find that our trends can be explained by the impact of $\Psi_{\mathrm{UV}}$ and $Z$ on the momentum from radiation pressure on dust (Section~\ref{sec:physics_trends}). Higher $\Psi_{\mathrm{UV}}$ leads to higher momentum per-unit stellar mass, whereas lower dust abundances ($Z$) reduces the important additional contribution from the dust-reprocessed infrared radiation pressure.
\end{enumerate}

\begin{acknowledgments}
We thank the referee for their insightful comments that have improved this manuscript. S.~H.~M would like to thank Piyush~Sharda, Laura Sommovigo and Chris Hayward for useful discussions. We acknowledge high-performance computing resources provided by the provided by the Simons Foundation as part of the CCA, and the Australian National Computational Infrastructure (grant~jh2) in the framework of the National Computational Merit Allocation Scheme and the ANU Merit Allocation Scheme. LL gratefully acknowledges the support of the Simons Foundation under grant 965367. AD has been partly supported by the Israel Science Foundation grant 861/20. B.B. and S.H.M. acknowledge the support of NASA grant No. 80NSSC20K0500 and NSF grant AST-2009679. B.B. also thanks the Alfred P. Sloan Foundation and the Packard Foundation for support. M.R.K. acknowledges support from the Australian Research Council through Laureate Fellowship FL220100020. This research has made use of NASA's Astrophysics Data System (ADS) Bibliographic Services. 
\end{acknowledgments}

%

\vspace{5mm}


\software{\texttt{SLUG} \citep{daSilva_2012,Krumholz_Slug}, \texttt{PETSc} \citep{PetscConf,PetscRef}, \texttt{NumPy} \citep{numpy}, \texttt{SciPy} \citep{scipy}, \texttt{Matplotlib} \citep{matplotlib}, \texttt{yt} \citep{yt}, \texttt{TurbGen} \citep{Federrath_2010,FederrathEtAl2022ascl}, \texttt{CFpack}.
          }



\appendix

\section{UV Luminosity of Stellar Populations}
\label{sec:appendix_slug}

In this appendix we present calculations of the UV output of stellar populations using the Stochastically Lighting Up Galaxies \citep[\texttt{SLUG};][]{daSilva_2012,Krumholz_Slug} stellar population synthesis code. We compute models of star cluster masses $M=\Mcloud=10^6 \, \Msun$ with increasingly top-heavy IMF prescriptions to motivate the parameter space of $\Psi$ we explore in our simulations. We use a \citet{Chabrier2005} functional form for the IMF of the stellar population with values for the mean and standard deviation of the lognormal part of the IMF ($M<1 \Msun$) as $0.2 \Msun$ and 0.55 dex, and a power law for $M>1 \Msun$ with slope $\alpha$, where $\alpha = -2.35$ is the value consistent with the \citet{Salpeter1955} IMF; the transition from the lognormal to the power-law occurs at $M = 1 \Msun$. We mimic a top-heavy IMF in two independent ways to test its impact on the UV luminosity of the stellar population -- i) changing the high-mass slope $\alpha$ while keeping the other parameters and functional forms the same, and ii) changing the upper limit of the stellar mass range of the IMF ($M_{\mathrm{max}}$) while fixing $\alpha = -2.35$, where $120 \Msun$ is the value for the standard fiducial IMF. We note that there are several other ways to achieve a top-heavy IMF, such as changing the peak of the IMF and/or the functional form itself, as predicted by several models \citep[e.g.,][]{Sharda_2022}. However, our intention with this calculation is simply to provide the reader a sense of the mapping between the IMF and the UV luminosities of the stellar populations.

\begin{figure}
    \centering
    \includegraphics[width=\textwidth]{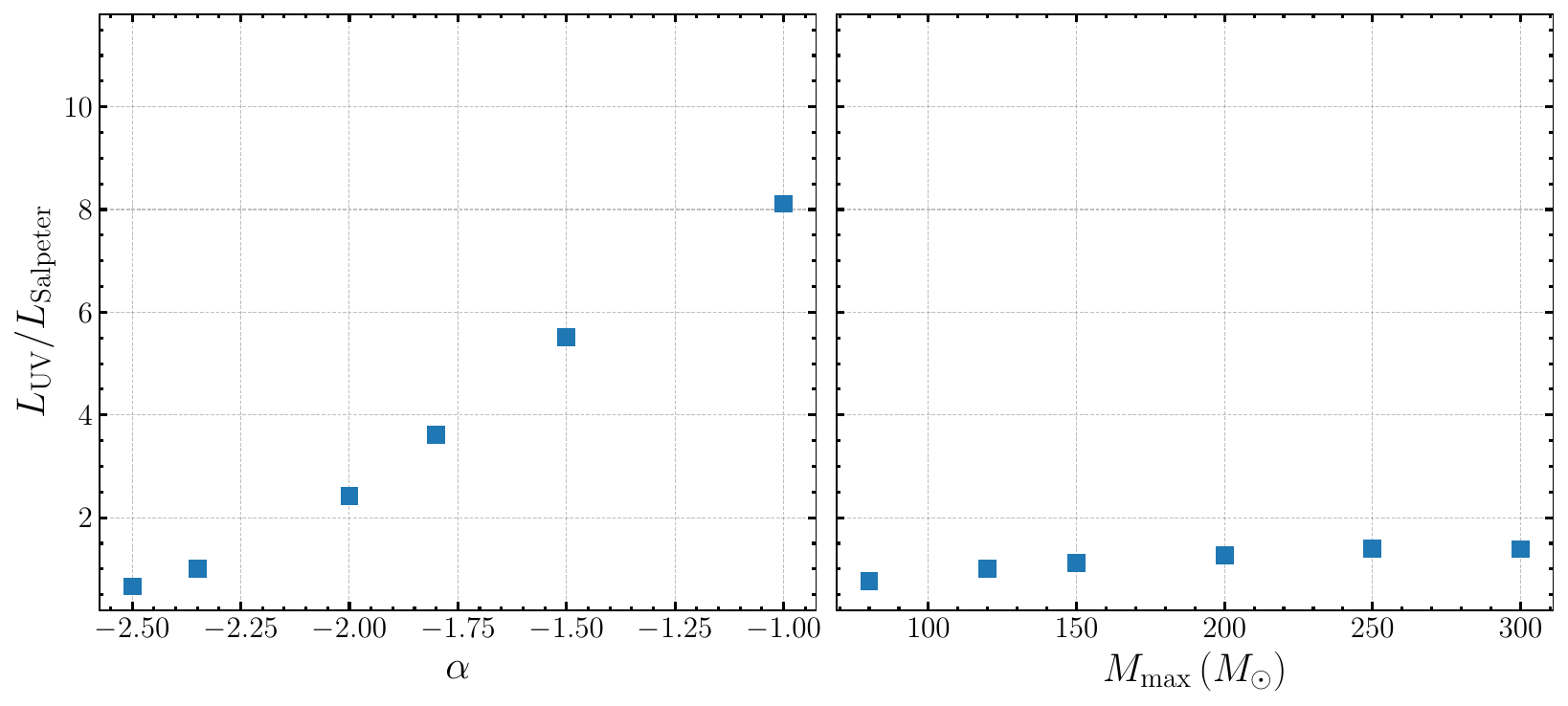}
    \caption{The UV luminosity per-unit mass of stars: i) for a given slope at the high-mass end of the IMF ($\alpha$; left panel), and ii) for a fixed Salpeter slope but different upper stellar mass limits ($M_{\mathrm{max}}$), both normalised by its counterpart for a standard Chabrier IMF -- i.e. a Salpeter slope (i.e. $\alpha = -2.35$) and an upper mass limit of 120$\Msun$. These values have been obtained with the \texttt{SLUG} stellar population synthesis code. The peak mass of the IMF and its shape are kept identical in this calculation -- only the high-mass slope and the upper mass limit are varied independently. We use this range of values ($\Psi_{\mathrm{UV}} \sim 1$--$10 \, \Psi_{\mathrm{Salpeter}}$) to motivate the parameter space we explore.}
    \label{fig:luminosity_alpha}
\end{figure}

We explore values of $-2.5 \leq \alpha \leq -1$ and $80 \Msun \leq M_{\mathrm{max}} \leq 300$. All our models use the \texttt{MIST} v1.0 isochrones \citep{Choi_2016} rotating at $40\%$ breakup, Solar metallicity, and \texttt{SLUG}'s \texttt{Starburst99} option for stellar atmospheres, which follows the approach described by \citet{Leitherer_1999} and implemented in the \texttt{Starburst99} stellar population synthesis code. 
We evolve our star cluster model for 5 Myr, and then compute the time-averaged UV ($h\nu \geq 6.6 \, \mathrm{eV}$) bolometric luminosity ($L_{\mathrm{UV}}$); the luminosities are relatively constant over this timescale after which it significantly drops due to the death of massive stars. In Figure~\ref{fig:luminosity_alpha} we report these values in units of the corresponding estimate for the \citealt{Salpeter1955} slope $\alpha = -2.35$ and $M_{\mathrm{max}} = 120 \Msun$. 

\section{Impact of Photoionization}
\label{sec:appendix_photoionization}

The simulations presented in this study do not include the photoionization of hydrogen, and the associated thermal pressure of photoionized gas that could act to regulate star formation. We argued that this omission does not impact our outcomes as the dynamical impact of the thermal pressure of photoionized gas is expected to be minor for these clouds with high escape velocities. We demonstrate this is the case by recomputing our metal-poor, highly top-heavy run (i.e. $Z=10^{-2} Z_{\odot}$, $\Psi_{\mathrm{UV}} = 10 \Psi_{\mathrm{Salpeter}}$) for $\Sigmacloud = 3 \times 10^4 \, \Msolpc$ with photoionization included. Our final normalized star formation efficiency $\epsf/\epsilon_{*,\mathrm{NoFB}}$ with and without photoionization are 92\% and 97\% respectively. This confirms that photoionization plays a very minor role in this regime, and thereby its omission does not affect our results. 

For completeness, we describe the additional parameters used for including the LyC band. We use the radiative luminosities for this run consistent with the values reported in Figure~\ref{fig:luminosity_alpha} for 
$\Psi_{\mathrm{UV}} = 10$. We use a cross section to ionizing photons for hydrogen corresponding to the Lyman edge $\sigma_{ion} = 6.3 \times 10^{-18} \, \mathrm{cm}^2$, and a constant (case-B) recombination coefficient $\alpha_{\mathrm{B}} = 2.6 \times 10^{-13} \, \mathrm{cm}^{3} \, \mathrm{s}^{-1}$. For simplicity, we also assume instantaneous thermal equilibrium for the ionized/neutral phases, by setting the temperature proportional to the ionization fraction such that fully ionized (neutral) gas has a temperature of $10^4 \, \mathrm{K}$ ($10 \, \mathrm{K}$) -- i.e. the two-temperature isothermal equation of state \citep[e.g.,][]{Gritschneder_2009,Kim_2018,Menon_2020}. We also include the associated radiation pressure on gas due to LyC absorption. 


\bibliography{references,federrath}{}
\bibliographystyle{aasjournal}



\end{document}